\documentclass[aps,
floatfix,superscriptaddress,a4paper]{revtex4}
\usepackage{graphicx,amsmath,bbm,mathrsfs,amssymb,psfrag}

\evensidemargin \oddsidemargin
\begin{document}

\author{Laure Gouba}

\affiliation{
The Abdus Salam International Centre for
Theoretical Physics (ICTP),
 Strada Costiera 11,
I-34151 Trieste Italy.
Email: lgouba@ictp.it}

\title{Beyond Coherent State Quantization}

\date{\today}

\begin{abstract}
We present an original approach to quantization based on operator-valued measure that generalizes the so-called Berezin-Klauder-Toeplitz quantization, and more generally coherent state quantization approches.
\end{abstract}

\maketitle

\section{Introduction}

In Physics, quantization is generally understood as a correspondence between a classical and a quantum theory and dequantization is the opposite process by which one starts with a quantum theory and arrives back at its classical counterpart. The processes of quantization and dequantization have evolved into mathematical theories, affecting the areas of group representation theory and symplectic geometry. Quantization is not a method for deriving quantum mechanics but a way to understand the deeper physical reality which underlies the structure of both classical and quantum mechanics and which unifies the two from a geometrical point of view. There is a certain mathematical richness in the various theories of quantization where the procedure does make sense. However, not every quantum system has a classical counterpart and then for such a system a quantization method does not make sense, moreover different quantum systems may reduce to the same classical theory.

Originally P. A. M. Dirac introduced the canonical quantization in his $1926$ doctoral thesis, {\it the method of classical analogy for quantization} \cite{pamd}. The canonical quantization or correspondence principle is an attempt to take a classical theory described by the phase space variables, let's say $p$ and $q$, and a Hamiltonian $H(q,p)$ to define or construct its corresponding quantum theory. The following simple technique for quantizing a classical system 
is used. Let $q^i, \: p_i, i= 1,2\ldots n$, be the canonical positions and momenta for a classical system with $n$ degrees of freedom. 
Their quantized counterparts $\hat q^i, \hat p_i$  are to be realized as operators on the Hilbert space $\mathcal{H} = L^2(\mathbb{R}^n, dx)$ by the prescription 
\begin{equation}
(\hat q^i\psi)(x) = x^i\psi(x);\quad 
(\hat p_i\psi)(x) = -i\hbar\frac{\partial}{\partial x^i}\psi(x);\quad i= 1,2,\ldots n,\quad x\in \mathbb{R}^n.
\end{equation}
This procedure is known as canonical quantization and is the basic procedure of quantization of a classical mechanics model 
\cite{dir,vneu,wey}. More general quantities, such as the Hamiltonians, become operators according to the rule 
\begin{equation}
H(p, q) \rightarrow \mathcal{H} =  H(\hat p , \hat q), 
\end{equation}
an expression that may have ordering ambiguities \cite{born, agar}.

In which canonical coordinates system does such a quantization procedure work?

\begin{enumerate}
\item According to Dirac replacing classical canonical coordinates by corresponding operators is found in practice to be successful only when applied with the dynamical coordinates and momenta referring to a cartesian system of axes and not to more general curvilinear coordinates.
\item Cartesian coordinates can only exist on a flat space.
\item The canonical quantization seems to depend on the choice of coordinates.
\item Beyond the ordering problem, one should keep in mind that 
$[Q,P] = i\hbar\,I_d$ holds true with self adjoint operators 
$Q,\;P$, only if both have continuous 
spectrum ($-\infty, +\infty$), and there is uniqueness of the solution, up to unitary equivalence (von Neumann).
\end{enumerate}
There are two attitudes that may be taken towards this apparent dependence of the procedure of the canonical quantization procedure on the choice of coordinates. 
The first view would be to acknowledge the cartesian character that is seemingly part of the procedure. The second view would be to regard it as provisional and seek 
to find a quantization formulation that eliminates this apparently unphysical feature of the current approaches. 
The aim of eliminating the dependence on cartesian coodinates in the standard approaches is no doubt one of the motivations for several procedures such as the geometric quantization \cite{geo1, geo2, geo3, geo4}, the path integral quantization \cite{fey}, the deformation quantization \cite{def1, def2, def3, def4, def5}, the Klauder-Berezin-Toeplitz quantization \cite{kbt1, kbt2, kbt3}.
 However, the canonical quantization is of the most famous and simplest procedure that is mostly accepted due to its experimental validations. In a recent review \cite{ali1} by S. T. Ali and
M. Engli$\check{s}$ various techniques of quantization conceived to make the transition from classical mechanics to quantum mechanics in the last five decades are presented.

The point of departure is always an analysis of the geometrical structure of either the classical phase space or the classical configuration space. There is no general theory of quantization presently available which is applicable in all cases, and indeed, often the techniques used to quantize has to be tailored to the problem in question.
As we have already said, canonical quantization generally requires the use of cartesian coordinates and not more general coordinates. So for a dynamical system without any constraints, its phase space  is assumed to be flat and admits a standard quantization of its canonical variables. 
In presence of constraints, all the variables of the system are not physical and the unphysical variables that may cause some little concerns at a classical level are highly unwelcome at the quantum level. It is then necessary to eliminate the unphysical variables and keep only the true physical degrees of freedom. The quantization of the physical degrees of freedom follows as in standard quantization. 

The study of constrained systems for the purpose of quantization was initiated by Dirac \cite{dirac1, dirac2, dirac3}, where he developed a method for treating singular systems and constraints. 
While Dirac's method and subsequent developments can cope with most models of interest, there are some problems in the sense that the quantization of the remaining degrees of freedom after one eliminates the unphysical variables may not be straightforward because the reduced (physical) phase space is non-euclidean in the sense that an obstruction has arisen where none existed before and that obstruction precludes the existence of self-adjoint canonical operators satisfying the canonical commutation relations.
Other approaches for quantizing constrained systems of classical theories have been subsequently 
developed \cite{faddee,fadjac,jan, jackiw,klau}.
Recently we have been interested in applying the Dirac's method in the study of the two-dimensional damped harmonic oscillator in the extended phase space \cite{lg} as some years ago, we have used the method to study nonperturbatively scalar and spinor abelian gauge theories in $1+1$ dimensions \cite{lau}. 
Inspired by Klauder's paper on coherent state quantization of constrained systems \cite{klaud}, we are interested in understanding a procedure of quantization beyond the coherent states quantization, generically called integral quantization mostly developed by J. P. Gazeau and al. \cite{gaz3,ali5,berg, jpgaz,gazmur}, in order to better 
study some of the constrained systems where the method is applicable. 
The paper is organized as follows. In section \ref{sec2} basic notions on coherent states are given. In section \ref{sec3}, starting from the Klauder-Berezin-Toeplitz quantization and the prime quantization, we reach the coherent state or frame quantization. In section \ref{sec4}, we start by introducing first the sea star algebra and then we  introduce the integral quantization. The perspectives are included in the later section.

\section{Basic notions on coherent states}\label{sec2}

Are coherent states the natural language of quantum theory? According to J. R. Klauder \cite{klau1}  the answer to the question is Yes!  In his paper he displayed various fundamental aspects of quantum theory from the perspective of a coherent state formulation.
The study of coherent states affect almost all branches of quantum physics: quantum optics, nuclear physics, atomic physics, solid -state physics, quantum electrodynamics (the infrared problem), quantization and dequantization problems, path integrals, quantum gravity, quantum information. Coherent-state methods have been proved convenient in the analysis of dual models of 
strings \cite{marinov}, and likewise they have been useful in the interacting boson model in nuclear physics, semiclassical calculations in chemistry have also made effective use of coherent state methods.

The first example of what is now called coherent states was discored by Schr\"odinger \cite{schr} as he was interested in studying quantum states which restore the classical behavior of the position operator of a quantum system in the Heisenberg picture,
\begin{equation}
Q(t) = e^{\frac{i}{\hbar}H t}Qe^{-\frac{i}{\hbar}H t} , 
\end{equation}
where $H = \frac{P^2}{2m} + V(Q)$ is the quantum Hamiltonian system.
Schr\"odinger understood by classical behavior that the expected value or average $\bar {q}(t)$ of the position operator $Q(t)$ in the desired state would obey the classical equation of motion 
\begin{equation}
m\ddot{\bar{q}}(t) + \frac{\partial \bar{V}}{\partial q} = 0.
\end{equation}
Schr\"odinger discovered the first example of coherent states 
pertaining to the harmonic oscillator $V(q) = \frac{1}{2}m^2\omega^2q^2$, known universally to physicists, and the prototype of every integrable model.
These states are parametrized by the complex number $z$, 
satisfying 
\begin{equation}
\langle z \vert Q(t)\vert z\rangle = 2 Q_0\vert z\vert 
\cos(\omega t- \varphi),
\end{equation}
$z = \vert z\vert e^{i\varphi}$ and $Q_0 = (\hbar/2 m\omega)^{\frac{1}{2}}$ is the fundamental quantum length, $\hbar$ the universal constant.

Using the generating function found in the classic book of Courant and Hilbert, Schr\"odinger realized that a Gaussian wave-function could be constructed from a particular superposition of the wave functions corresponding to the discrete eigenvalues of the harmonic oscillator. Further these new states followed the classical motion.
At this time the probability-amplitude nature of the wave function was not yet known, so the complex nature of the wave function bothered Schr\"odinger. He wondered if, perhaps, it was only the real part of the wave function that is physical. At that time the uncertainty relations had yet to be discovered, but from the point of view that most closely resembles the modern minimum uncertainty method, Schr\"odinger had discovered the coherent states.

The notion of coherent states is rooted in quantum physics and its relation to classical physics and the term coherent was introduced by R. J. Glauber in $1963$ in the field of quantum optics, and it is for a special property in quantum optics that the name coherent states was originally chosen, for instance from the current language of quantum optic there are expressions like coherent radiation, sources emitting coherently, $\ldots$ The importance of coherent states became widely recognized during the $1960$'s due to the works of Glauber, Klauder and 
Sudarshan \cite{glau1,glau2,jklau1,jklau2,sudar1,sudar2}.
Although the name {\it coherent states} is applied to a wide class of objects, in every case the set of states referes to vectors in a Hilbert space $\mathcal{H}$ (finite or countably infinite dimensional). 
We denote the states in question, in Dirac notation, by $\vert l\rangle$, where $l$  is an element (in general multidimensional) of an appropriate label space $\mathcal{D}$ endowed with a notation of continuity (topology). More specifically, an appropriate label space is one for which every finite dimensional subspace is locally euclidean.
There are in essence just two properties that all coherent states share in common:
\begin{enumerate}
\item Continuity: the vector $\vert l\rangle$ is a strongly continuous function of the label $l$.
\item  Resolution of unity (completeness): there exists a positive measure $\delta l $ on $\mathcal{D}$ such that the unit operator $I_d$ admits the "resolution of unity" $I_d = \int\vert l\rangle\langle l\vert \delta l$ when integrated over $\mathcal{D}$.
\end{enumerate}

The Schr\"odinger-Klauder-Glauber-Sudarshan coherent states, denoted here $\vert z\rangle$, also called canonical coherent states or standard coherent states satisfy the following properties:
\begin{enumerate}
\item the states $\vert z\rangle$ saturate the Heisenberg inequality
$\langle \Delta Q\rangle_z\langle \Delta P\rangle_z = \frac{1}{2}\hbar$;
\item the states $\vert z\rangle $ are eigenvectors of the annihilation 
operator, with eigenvalue $z$, $a\vert z\rangle = z\vert z\rangle$, $z\in \mathbb{C}$;
\item the states $\vert z\rangle $ are obtained from the ground state
$\vert 0 \rangle$ of the harmonic oscillator by a unitary action 
of the Weyl Heisenberg group,
\begin{equation}
\vert z\rangle  = e^{za^\dagger -\bar{z} a}\vert 0\rangle;
\end{equation}
\item the coherent state $\{\vert z\rangle \}$ constitute an overcomplete family of vectors in the Hilbert space of the states of the Harmonic oscillator. This property is encoded in the following resolution of the identity 
\begin{equation}
I_d = \frac{1}{\pi}\int_{\mathbb{C}}\vert z\rangle\langle z\vert d Re(z)d Imz.
\end{equation}
\end{enumerate}
These four properties are, to various extents, the basis 
of the many generalizations of the canonical notion of coherent 
state, illustrated by the family $\{\vert z \rangle\}$.
The unique qualification of the coherent states to allow for a 
classical {\it reading} in a host of quantum situations results from these four properties.

\section{Coherent states quantization}\label{sec3}

Quantization procedure is an important aspect of coherent states, for instance standard coherent states offer a classical-like representation of the evolution of quantum observables.
In coherent states quantization, we are interested in quantization of sets, more precisely measure spaces through coherent states.
For the normalized coherent states $\vert z \rangle$, the resolution of the unity 
\begin{equation}
\frac{1}{\pi}\int_\mathbb{C}\vert z\rangle\langle z\vert d^2z = I_d
\end{equation}
is a crucial property for our purpose in setting the bridge between the classical and the quantum worlds.

\subsection{Klauder-Berezin-Toeplitz quantization}

The Klauder-Berezin coherent state  quantization also named anti-Wick quantization or Toeplitz quantization by many authors, consists in associating with any classical operator $f$ that is a function of phase space variable $(q,p)$ or equivalently of $(z,\bar z)$, the operator valued integral 
\begin{equation}
\frac{1}{\pi}\int_\mathbb{C} f(z,\bar z)\vert z\rangle\langle z\vert d^2 z = A_f.
\end{equation}
The function $f$ is usually supposed to be smooth, but we will not retain in the sequel this too restrictive attribute.
The resulting operator $A_f$, if it exists, at least in a weak sense, acts on the Hilbert space $\mathcal{H}$ of quantum states for which the set of Fock (or number ) states $\vert n\rangle$ is an orthogonal basis. It is worthwhile being more explicit about what we mean by {\it weak sense}: the integral 
\begin{equation}\label{ii}
\langle \psi \vert A_f\vert \psi\rangle = \int_{\mathbb{C}}f(z,\bar z)
\vert\langle \psi\vert z\rangle\vert^2\frac{d^2 z}{\pi}
\end{equation}
should be finite for any $\vert \psi\rangle$ in some dense subset in $\mathcal{H}$. 
One should note that if $\vert \psi\rangle$ is normalized then the integral (\ref{ii}) represents the mean value of the function $f$ with respect to the $\psi$ -dependent probability distribution $z \rightarrow \vert \langle \psi\vert z\rangle\vert^2$ on the phase space. More mathematical rigor is necessary here, and we will adopt the following acceptance criteria  for a function (or distribution) to belong to the class of quantizable classical observables. A function $f: \mathbb{C} \rightarrow \mathbb{C},\quad z\mapsto f(z,\bar z)$ (more generally a distribution $T\in \mathcal{D}'(\mathbb{R}^2)$ ) is a coherent state quantizable classical observable along the map $f\mapsto A_f$ defined by $ A_f = \frac{1}{\pi}\int_\mathbb{C} f(z,\bar z)\vert z\rangle\langle z\vert d^2 z$ ( more generally  $T\mapsto A_T$) :
\begin{enumerate}
\item if the map $z\mapsto \langle z\vert A_f\vert z\rangle$ (or $z\mapsto \langle z\vert  A_T\vert z\rangle $), $z = \frac{1}{\sqrt 2}(q + ip)\equiv (q,p)$, is a smooth ( $\in C^\infty $ ) function with respect to the $(q,p)$ coordinates of the phase space,
\item and if we restore the dependence on $\hbar$ through $ z\rightarrow \frac{z}{\sqrt \hbar}$, we must get the right semiclassical limit, which mean that $\langle \frac{z}{\sqrt \hbar}\vert A_f\vert \frac{z}{\sqrt\hbar}\rangle\rightarrow 
f(\frac{z}{\sqrt\hbar},\frac{\bar z}{\sqrt\hbar})$,
\end{enumerate}
the same asymptotic behavior must hold in a distributional sense if we are quantizing distributions.

\subsection{Ordering problem in quantum mechanics: prime quantization}

Let assume we have a classical observable, precisely a real-valued function $f$ of the position variable $q$ and the momentum variable $p$.
We assume that the configuration space of the system is $\mathbb{R}^1$ means that it moves on the phase space 
$T^\star (\mathbb{R}^1)\sim \mathbb{R}^2$.
If $f$ admits a Taylor expansion, one has 
$f(q,p) =\sum_{m,n =0}^\infty C_{mn}q^m p^n$, 
where $C_{mn}$ are appropriate coefficients which could possibly be zero for $m,n > N$ for some integer $N$.
In a quantized theory, the variables $q,p$ are replaced by essentially self-adjoint unbounded operators $Q$ and $P$, respectively, on a (separable) Hilbert space $\mathcal{H}$, satisfying the canonical commutation relations (over a common dense domain $\mathcal{D}$, $[Q,P] = i,\; (\hbar = 1)$. What the self-adjoint operator $F$ on $\mathcal{H}$, corresponding to the classical observable $f$ should be? It seems plausible to quantize $f$ via $F = \sum_{m,n =0}^\infty C_{mn}Q^m P^n$. Although the infinite sum of unbounded operators appearing in 
$F = \sum_{m,n =0}^\infty C_{mn}Q^m P^n$ has a meaning and even if $F$ was a self-adjoint operator on a dense domain $\mathcal{D}_F\subset \mathcal{H}$, one has to decide on a certain ordering of the noncommutating operators $Q$ and $P$, when their products appear in $F = \sum_{m,n =0}^\infty C_{mn}Q^m P^n$.
For instance, it is a matter of some arbitrariness as to whether the classical observable $qp^2$ is to be replaced by the quantum operator $\frac{1}{2}[QP^2 + P^2 Q]$ or $\frac{1}{4}[QP^2 + 2PQP + P^2Q]$ or some other self-adjoint combinations.

Some of the combinations are mathematically equivalent, however, in our opinion, not enough physical justification is given for choosing one ordering over another.
We introduce the  prime quantization, that is a quantization technique developped by H. D. Doebner \cite{ali2, ali3, doeb}. This technique sheds light on the time honoured ordering problem of quantum mechanics and has connections to coherent state quantization and Klauder-Berezin-Toeplitz quantization.
The method starts out on the phase space $\Gamma$ of the classical system and exploits the functional analytical rather than geometric structures of the system.
One starts with the symplectic manifold $(\Gamma, \Omega)$ and the Hilbert space $\mathcal{H} = L^2(\Gamma,\Omega)$. The next step is to look for reproducing kernel subspaces of $\mathcal{H}$. Let $\mathcal{H}_K$ a subspace of $\mathcal{H}$, on 
which the evaluation maps
\begin{equation}
E_\gamma: \mathcal{H}_K\rightarrow C, \quad E_\gamma(\psi) =\psi(\gamma), \quad \gamma\in \Gamma, \psi\in \mathcal{H}_K
\end{equation} are continuous.
Then, $K(\gamma,\gamma') = E_\gamma E_{\gamma'}^\star$ is a reproducing kernel.
For any Borel set $\Delta \subset \Gamma$, the operator
\begin{equation}
a_K(\Delta) = \int_\Delta E^\star_{\gamma'} E_\gamma d\mu(\gamma),
\end{equation}
on $\mathcal{H}_K$ is positive.

A classical observable $f: \Gamma\rightarrow \mathbb{R}$
is then quantized by the prescription, 
\begin{equation}
Q(f) = \int_\Gamma f(\gamma)E^\star_{\gamma'} E_\gamma d\mu(\gamma).
\end{equation}
One has to prove the asymptotic validity of
\begin{equation}
[Q(f), Q(g)] \sim i\hbar Q(\{f,g\}).
\end{equation}
The challenge in this method is to find the appropriate reproducing 
kernel Hilbert space $\mathcal{H}_K$, which is reflective of the physical problem.

\subsection{Coherent states or frame quantization}

We consider the Hilbert space $L^2_K(X,\mu)(K = \mathbb{R} $ or $K = \mathbb{C}$) of square-integrable 
real or complex functions $f(x)$ on the observation set $X: \int_X\vert f(x)\vert^2\mu(dx) <\infty$. 
A classical observable is a function $f(x)$ on $X$ having specific properties with respect to some 
supplementary structure allocated to $X$, such as topology, geometry, or something else. 

Given a frame{\footnote{more about the concept of frame can be read in references \cite{ali4,gaz1}}} $\{\vert x\rangle, x\in X \}$, a function $ f: x (\in X) \mapsto f(x) (\in K)$, possibly undertood in a distribution sense, is a coherent state quantizable classical observable along the map $f\mapsto A_f = \int_X f(x)\vert x\rangle\langle x\vert \nu(dx)$, if the map $x (\in X) \mapsto \langle x\vert A_f\vert x\rangle$ is a regular function with respect to some additional structures allocated to $X$.
We must get the right classical limit, which means that $\langle x\vert A_f \vert x\rangle = f(x)$ as a certain parameter goes to 0.

A quantization of the observation set $X$ is in one-to-one correspondence with the choice of a frame in the sense of $\int_X\vert x\rangle\langle x \vert \nu(dx) =  I_d$ where $\nu(dx) = \mathcal{N}(x)\mu(dx)$. This term of frame is more appropriate for designating the total family $\{\vert x\rangle\}_{x\in X}$.
This frame can be discrete or continuous, depending on the topology additionnaly allocated to the set $X$, and it can be overcompleted.
The validity of a precise frame choice is determined by comparing spectral characteristics of quantum observables $A_f$ with experimental data.
A quantization method associated with a specific frame is intrinsically limited to all those classical observables for which the expansion $A_f =  \int_Xf(x)\vert x\rangle\langle x\vert \nu(dx)$ is mathematically justified within the theory of operators in Hilbert space (weak convergence).

The coherent state quantization in this context requires a minimal significant structure on $X$, namely, the existence of a measure $\mu(dx)$, together with a $\sigma$-algebra  of measurable subsets, and some additional structure to be defined depending on the context.
The construction of the Hilbert space $\mathcal{H}$ is equivalent to the choice of a class of eligible quantum states, together with a technical condition of continuity. A correspondence between classical and quantum observables is then provided through a suitable generalization of the standard coherent states.

\section{Beyond coherent state quantization: integral quantization}\label{sec4}

Let us brieffly recall the following. 
We consider an orthonormal basis (or frame) of the Euclidean plane $\mathbb{R}^2$ 
defined by the two vectors (in Dirac ket notations)
\begin{equation}
\vec{i} \equiv \vert 0\rangle \equiv 
\left(\begin{array}{cc}
1\\
0
\end{array}
\right)\quad \textrm{and}\quad 
\vec{j} \equiv \vert \frac{\pi}{2}\rangle \equiv 
\left(\begin{array}{cc}
0\\
1
\end{array}
\right),
\end{equation}
and $\vert \theta\rangle$ denotes the unit vector with polar angle $\theta \in [0,\;2\pi)$.
This basis or frame is such that $\langle 0\vert 0 \rangle = 1 = 
\langle \frac{\pi}{2}\vert \frac{\pi}{2} \rangle $ and $\langle 0\vert \frac{\pi}{2}\rangle= 0$. 
The sum of their corresponding orthogonal projectors resolves the unit 
$I_d = \vert 0\rangle\langle 0\vert + \vert \frac{\pi}{2}\rangle\langle \frac{\pi}{2}\vert$ , 
that is equivalent to 
\begin{equation}
\left(
\begin{array}{cc}
 1 & 0\\ 0 & 1
\end{array}\right)  =
\left(
\begin{array}{cc}
 1 & 0\\ 0 & 0
\end{array}\right) +
\left(
\begin{array}{cc}
 0 & 0\\ 0 & 1
\end{array}\right).
\end{equation}
To the unit vector $\vert \theta\rangle = \cos\theta\vert 0\rangle + 
\sin\theta\vert\frac{\pi}{2}\rangle$ corresponds the orthogonal projector 
$P_\theta$ given by 
$P_\theta  = \vert \theta\rangle\langle \theta\vert$  and 
\begin{equation}
\vert \theta \rangle \langle \theta\vert = 
\left(\begin{array}{cc}
\cos\theta\\\sin\theta\end{array}\right)
\left( \cos\theta\:\: \sin\theta\right)
= \left(
\begin{array}{cc}
\cos^2\theta & \cos\theta\sin\theta\\
\cos\theta\sin\theta & \sin^2\theta
\end{array}\right)
\end{equation}
$P_\theta = \mathcal{R}(\theta)\vert\theta\rangle\langle\theta\vert\mathcal{R}(-\theta)$, where 
\begin{equation}
\mathcal{R}(\theta) = \left(
\begin{array}{cc}
\cos\theta  & -\sin\theta\\
\sin\theta & \cos\theta
\end{array}
\right)\in SO(2).
\end{equation}

\subsection{Sea star algebra}

The five-fold symmetry of sea star (starfish) motivated the development in a comprehensive way, the quantization of functions on a set with five elements which yields in particular the notion of a quantum angle.
The purpose of this introductory model is to enhance our intuition that quantization can be viewed as the analysis of a set from the point of view of a family of coherent states.

\begin{figure}[h]
\centering
  \includegraphics[width= 6cm]{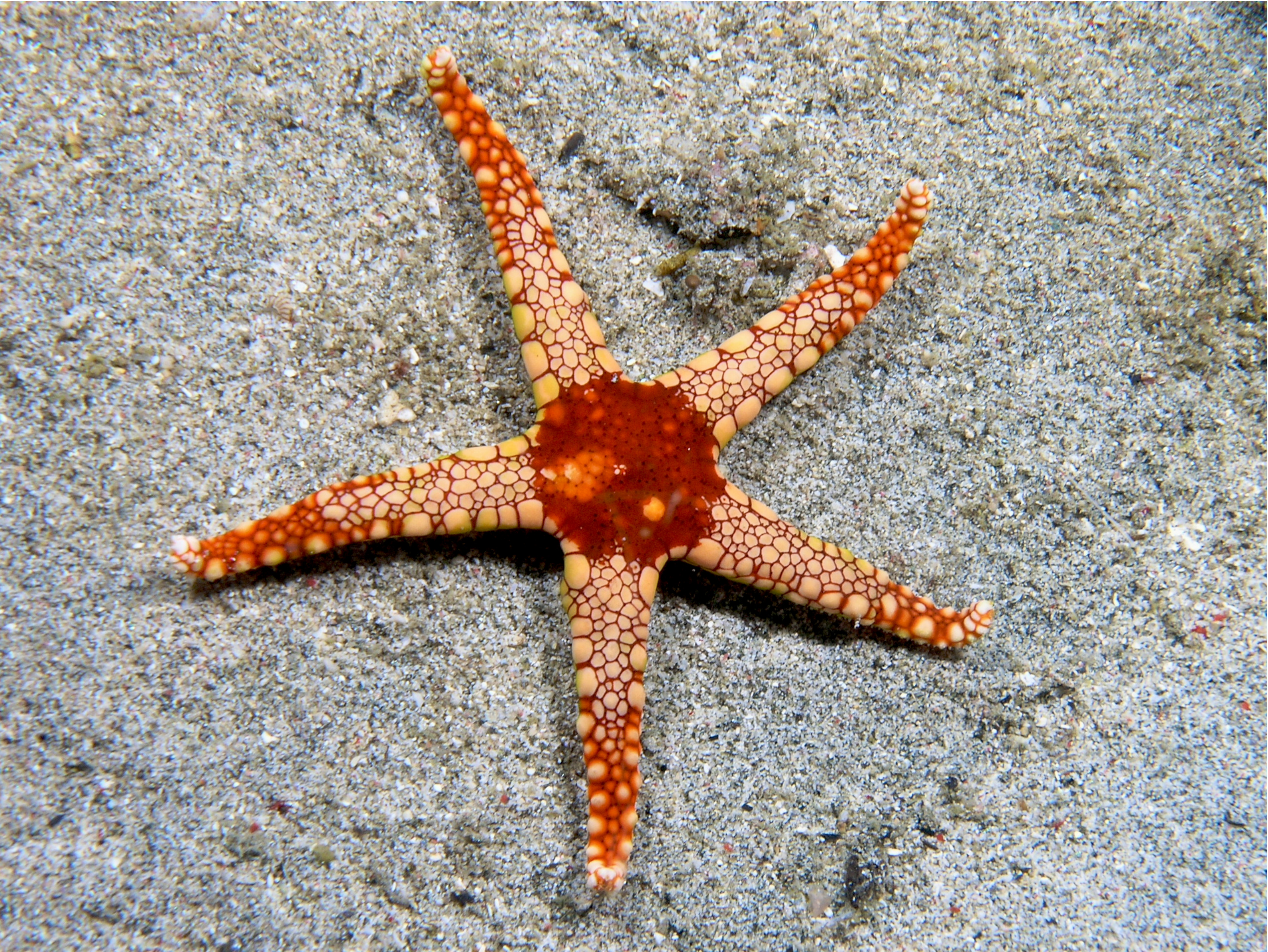}
  \caption{A sea star as a five-fold frame for the plane}
  \label{fig:seastar}
\end{figure}

\begin{enumerate}
\item sea stars have five arms, or rays, connected to a small round body;
\item sea stars detect light with five purple eyespots at the end of each arm;
\item sea stars typically show pentameral symmetry.
\end{enumerate}

How the ocean floor is "five-fold orientationally ' explored by a sea star?

A sea star possibly do this in a noncommutative way through the pentagonal set of unit vectors (the arms) 
forming a five-fold frame in the plane.
The unit vectors (arms) are determined by $\vert \frac{2n\pi}{5}\rangle = 
\mathcal{R}(\frac{2n\pi}{5})\vert 0\rangle \equiv$ "coherent" state $n = 0,1,2,3,4$ mod 5.
We get the resolution of the identity:

\begin{equation}
\frac{2}{5}\sum_{n=0}^4\vert \frac{2\pi n}{5}\rangle\langle 
\frac{2\pi n}{5}\vert =  \left(\begin{array}{cc} 1 & 0\\
0 & 1\end{array}\right)\equiv I_d.
\end{equation}

The property holds for any regular N-fold polygon in the plane
\begin{equation}
\frac{2}{N}\sum_{n=0}^{N-1}\vert \frac{2\pi n}{N}\rangle\langle 
\frac{2\pi n}{N}\vert =  \left(\begin{array}{cc} 1 & 0\\
0 & 1\end{array}\right).
\end{equation}

Let's make more precise the set $X$ explored by the sea star, determine a probabilistic construction of the 
frame by using the Hilbert space structure of the plane and
an exact way of exploring $X$ through functions on this finite set. The sea star senses its possible orientations via the five angles $\frac{2\pi n}{5}$ and $X = \{0,1,2,3,4\}$ the set of orientations. It is equipped with discrete measure with uniform weight 
\begin{equation}
\int_x f(x)d\mu(x) = \frac{2}{5}\sum_{n=0}^4 f(n).
\end{equation}
Let us choose two orthonormal elements, $\phi_0(n) =\cos\frac{2\pi n}{5} $ and 
$\phi_1(n) = \sin \frac{2\pi n}{5}$ in the Hilbert space $L^2(X,\mu)$, and build the five unit vectors in the real two-dimensional Hilbert space, that is the Euclidean plane $\mathbb{R}^2  = \mathcal{H}$, with the usual orthonormal basis $\vert 0\rangle,\; \vert \frac{\pi}{2}\rangle$:
\begin{equation}
n\in X,\quad n\mapsto \vert n\rangle \equiv \vert \frac{2\pi n}{5} \rangle 
= \phi_0(n)\vert 0\rangle +\phi_1(n)\vert \frac{\pi}{2}\rangle.
\end{equation}
The operators $P_n = \vert\frac{2\pi n}{5}\rangle \langle \frac{2\pi n}{5}\vert $ acts on 
$\mathcal{H} = \mathbb{R}^2$. Given a $n_0 \in \{0,1,2,3,4\}$, one derives the probability distribution 
on $X = \{0,1,2,3,4\}$ 
\begin{equation}
\textrm{tr}(P_{n_0} P_n) = \vert \langle \frac{2\pi n_0}{5}\vert \frac{2\pi n}{5}\rangle\vert^2 = \cos^2 \left(\frac{2\pi(n_0 -n)}{5}\right).
\end{equation}
The five unit vectors $\vert n\rangle\equiv \vert \frac{2\pi n}{5}\rangle$ resolve the identity in $\mathcal{H}$. They form a finite unit frame for analysing the complex-valued functions $n\mapsto f(n)$ on $X$ through what we call a coherent state (CS) quantization:
\begin{equation}
f(n)\mapsto \int_X\vert x\rangle \langle x\vert f(x) d\mu(x) = \frac{2}{5}\sum_{n=0}^4 f(n)\vert\frac{2n\pi}{5}\rangle\langle \frac{2n\pi}{5}\vert \equiv A_f.
\end{equation}
If we had chosen instead as a finite frame the orthonormal basis 
$\vert 0\rangle, \vert \frac{\pi}{2} \rangle$, in
$\mathbb{R}^2$, we would have obtained the trivial commutative quantization: 
\begin{equation}
(f(0), f(1))\mapsto A_f= \textrm{diag}(f(0), f(1)).
\end{equation}
This map should be analysed more precisely through spectral values of the $2\times 2$ symmetric matrix $A_f$, and also through coherent states mean values of $A_f$: 
\begin{equation}
n\mapsto \check{f}(n) := \langle \frac{2n\pi}{5}\vert A_f\vert \frac{2 n\pi}{5}\rangle 
= \frac{2}{5}\sum_{m=0}^4 f(m)\cos^2\frac{2(n-m)\pi}{5},\; n = 0,1,2,3,4 .
\end{equation}

\subsection{Integral quantization}
An original approach to quantization based on operator valued measures is named generically integral quantization. The approach generalize the coherent state quantization. The probabilistic aspects appearing is highlighted at each stage of the quantization procedure. To be more mathematically precise and still remaining at a basic level, quantization is a linear map $\mathcal{Q}:\mathcal{C}(X)\rightarrow \mathcal{A}(\mathcal{H})$ from a vector space $\mathcal{C}(X)$ of complex valued function $f(x)$ on a set $X$ to a vector space $\mathcal{A}(\mathcal{H})$ of linear operator $\mathcal{Q}(f)\equiv A_f$ in some complex Hilbert space $\mathcal{H}$, 
such that 
\begin{enumerate}  
\item to the function $ f = 1$ there corresponds the identity operator $I_d$ on $\mathcal{H}$;
\item to a real function $f\in \mathcal{C}(X)$ there corresponds 
a  (an essentially) self adjoint operator $A_f$ in $\mathcal{H}$.
\end{enumerate}
The above conditions may be easily fulfilled if one uses the resources offered by the pair (measure, integration).
Let $(X,\nu) $ be a measure space. Let $\mathcal{H}$ be a complex Hilbert space and $x\in X,\; x\mapsto M(x)\in \mathcal{L}(\mathcal{H})$ an $X$-labelled family of bounded operators on $\mathcal{H}$ resolving the identity:
\begin{equation}
\int_X M(x) d\nu(x) = I_d,
\end{equation}
provided that the equality is valid in a weak sense, which implies $\nu$ integrability for the family $M(x)$. If the operators $M(x)$ are positive and have unit trace, they will be preferentially denoted by $M(x) = \rho(x)$ in order to comply with the usual notation for a density matrix in quantum mechanics.
The corresponding quantization of complex-valued functions $f(x)$ on $X$ is then defined by the linear map:
\begin{equation}
 f \mapsto A_f =  \int_X M(x)f(x)d\nu (x).
\end{equation}

This operator-valued integral is again understood in the weak sense, 
as the sesquilinear form, 
\begin{equation}
B_f(\psi_1,\psi_2) = \int_X \langle \psi_1\vert M(x)\vert\psi_2\rangle f(x)d\nu(x).
\end{equation}
The form $B_f$ is assumed to be defined on a dense subspace of $\mathcal{H}$.
 If $f$ is real and at least semi-bounded, the Friedrichs extension of $B_f$ univocally defines 
 a self-adjoint operator.
If $f$ is not semi-bounded, there is no natural choice of a self-adjoint operator associated with $B_f$, in this case, we can consider directly the symmetric operator $A_f$, enabling us to obtain a self-adjoint extension (unique for particular operators).

The situation when the operators $M(x)$ are negative has not been discussed here and may be an interesting aspect to check. The spectral properties of the operators $A_f$ may be analysed. An interesting aspect of this integral quantization method is the 
possibility to quantize constraints.

\noindent{\bf Acknowledgement }: {This paper is a slightly enlarged version of the talk given at the XXVIth International Conference on Integrable Systems and Quantum symmetries (ISQS-26), Prague, Czech Republic, 08 -12 July 2019. L. Gouba would like to acknowledge support from ICTP/Trieste and to thank the organizing committee of the ISQS-26.}

\vspace{0,5cm}

\end{document}